\documentclass[final,5p,times,twocolumn]{elsarticle} 
\usepackage{epsfig}  
\usepackage{amssymb,amsmath}
\usepackage{tabularx}
\usepackage[table,xcdraw]{xcolor}
\usepackage{natbib}
\usepackage[greek,english]{babel}
\usepackage{subcaption}
\usepackage{float}
\usepackage[colorlinks=true,linkcolor=blue,urlcolor=blue]{hyperref}
\usepackage{xcolor}
\usepackage{ulem}
\usepackage{svg}

\journal{Physica A: Statistical Mechanics and its Applications}
\makeatletter
\def\ps@pprintTitle{%
 \let\@oddhead\@empty
 \let\@evenhead\@empty
 \let\@oddfoot\@empty
 \let\@evenfoot\@oddfoot
}
\makeatother
\begin{document}

\begin{frontmatter} 
\title{Wealth Inequality in Agent-Based Economies: \\ The Dominant Role of Social Protection over Growth}






\author[gaston]{Gastón Villafañe}
\ead{gaston.villafane@ib.edu.ar}

\author[lauti]{Lautaro Giordano}
\ead{lautaro.giordano@ib.edu.ar}

\author[fabi]{María Fabiana Laguna\corref{cor1}}
\ead{lagunaf@cab.cnea.gov.ar}
\cortext[cor1]{Corresponding author}

\address[gaston]{Instituto Balseiro, Universidad Nacional de Cuyo. R8402AGP Bariloche, Argentina.}

\address[lauti]{Statistical and Interdisciplinary Physics Group, Centro Atómico Bariloche (CNEA) and CONICET. Instituto Balseiro, Universidad Nacional de Cuyo. R8402AGP Bariloche, Argentina.}

\address[fabi]{Statistical and Interdisciplinary Physics Group, Centro Atómico Bariloche (CNEA) and CONICET. R8402AGP Bariloche, Argentina.}

\begin{abstract}
Persistent wealth inequality, where a small fraction of the population accumulates most resources while the majority remains economically vulnerable, is a widespread phenomenon. We investigate its underlying mechanisms using an agent-based Yard-Sale model that incorporates two complementary features: transaction rules that favor poorer agents, representing social protection policies, and an economic growth process with explicit wealth redistribution. 
Our results reveal that social protection plays a dominant role in reducing inequality, while redistribution primarily serves to reintegrate excluded agents. These findings suggest that social protection policies, that is, targeted mechanisms favoring economically vulnerable agents, may have a substantially greater impact on reducing inequality than redistribution driven solely by economic growth. We also find that both the shape of the wealth distributions and the resulting inequality levels are strongly influenced by the underlying distribution of individual risk, highlighting the importance of considering agent heterogeneity when modeling economic dynamics.
\end{abstract}

\begin{keyword}
Econophysics, Wealth distribution, Kinetic wealth exchange models, Inequality, Yard-Sale model 
\end{keyword}

\date{\today}
\end{frontmatter}

\section{Introduction}

The capitalist economic system is fundamentally driven by the accumulation of capital and the pursuit of profit. Within this framework, goods and services are typically produced and exchanged to maximize their market value, emphasizing the predominance of exchange value over use value. Economic agents are assumed to act rationally to increase their personal wealth, often prioritizing individual gain over collective welfare. These foundational assumptions, namely rational behavior of agents and profit maximization, have long served as the foundation of traditional economic models~\cite{samuelson1947}, and remain central to capitalist economies, which prevail in most countries today and have shaped socioeconomic interactions for centuries. 

In recent decades, the increasing availability of large-scale economic data has motivated researchers to explore the statistical structure of wealth and income distributions. Empirical studies have highlighted persistent and pronounced inequality~\cite{piketty2014}, prompting the adoption of tools from statistical physics to model these phenomena and uncover their underlying mechanisms. 
This effort has given rise to the interdisciplinary field of Econophysics, which applies concepts and methods from statistical physics to the study of complex economic systems~\cite{castellano2009,chatterjee2007}. Within this framework, several works have shown that the accumulation-driven logic of capitalism and the rules of wealth exchange can inherently lead to sustained inequality~\cite{yakovenko2009,boghosian2017}.

A striking feature of human societies is the concentration of wealth: a small fraction of the population holds a disproportionate share of resources while the majority remains in poverty. The persistence of this pattern across time and cultures suggests the presence of intrinsic mechanisms behind wealth condensation. One such mechanism, proposed by Boghosian~\cite{boghosian2019}, emphasizes asymmetries in wealth exchange. While ideal transactions are assumed to be neutral, real-world exchanges are rarely balanced. Strategic behavior, market fluctuations, and imperfect information introduce biases that disproportionately favor wealthier agents, who are more resilient to loss and better positioned to benefit from such asymmetries~\cite{chatterjee2007,li2021}. This dynamic contributes to a self-reinforcing process of inequality~\cite{boghosian2017,boghosian2020}.

To investigate these dynamics, kinetic exchange models such as the Yard-Sale model have been widely employed~\cite{boghosian2017,hayes2002follow,giordano2025}. In this model, agents trade an amount proportional to the minimum of their current wealth, and the winner is chosen at random. Although the rule is symmetric, both numerical and analytical studies show that the system evolves toward an absorbing state in which wealth condenses into the hands of a single agent or small elite~\cite{chatterjee2007,yakovenko2009,boghosian2017,boghosian2020}. This paradoxical outcome underscores how inequality can emerge even under apparently fair rules.

In this work, we introduce a generalized Yard-Sale model that combines two mechanisms aimed at mitigating inequality, drawing specifically on the implementations presented in~\cite{nener2021} and~\cite{liu2021} as reference points: a social protection factor that favors poorer agents in transactions, and a process of wealth injection coupled with redistribution.
Rather than studying each mechanism in isolation, we analyze their joint effect within a unified framework. Our goal is to characterize how these mechanisms interact and influence the system’s dynamics and stationary wealth distribution.
Furthermore, while~\cite{nener2021} assigns agents heterogeneous risk values drawn from a uniform distribution,~\cite{liu2021} assumes a common risk parameter for all. Inspired by these contrasting choices, we incorporate both types of risk distributions into our model. This allows us to assess the role of agent heterogeneity in shaping the effectiveness of social protection and redistribution policies.

The rest of the paper is organized as follows. In the next section, we define the model in detail, including the exchange rules, the implementation of social protection, and the wealth injection mechanism. We then present our numerical results, beginning with the homogeneous risk case and later extending to heterogeneous scenarios. Finally, we summarize our main findings and discuss their implications.


\section{Methodology}
\subsection*{Yard-Sale model}

The model consists of an ensemble of $N$ interacting agents, each provided with wealth $w_i$ and a fixed risk value $r_i \in [0,1]$\footnote{Also referred as the complement of the \textit{risk aversion}~($\beta_i = 1-r_i$) in the literature.}, representing the fraction of wealth that agents are willing to risk in an exchange. Given two agents $i$ and $j$, the amount of money traded in an exchange is given by the Yard-Sale rule
\begin{equation} \label{eq:ysrule}
\Delta w_{ij} = \min(r_i w_i, r_j w_j).   
\end{equation}

The wealth of the interacting agents is then updated according to:
\begin{equation}
\begin{split}
w_i^* &= w_i + (2\eta_{ij} - 1)\Delta w_{ij}, \\
w_j^* &= w_j - (2\eta_{ij} - 1)\Delta w_{ij},
\end{split}
\end{equation}
where $\eta_{ij}$ is a Bernoulli random variable that takes the value $1$ with probability $p$. Regardless of the outcome ($\eta_{ij} = 0$ or $\eta_{ij} = 1$), the exchange rule conserves wealth, so the total wealth of the system $W = \sum_{i=1}^N w_i$ remains constant. The initial wealths of the agents are randomly drawn from a uniform distribution and then normalized so that the total wealth at time $t = 0$ is $W(0) = N$, ensuring a mean wealth per agent of $\langle w \rangle = 1$.

Agents whose wealth falls below a minimum threshold, \(w_{\text{min}}\), are considered bankrupt and no longer participate in the dynamics. We adopted \(w_{\text{min}} =  10^{-17}\) for all simulations, a value sufficiently small to ensure numerical stability while having no appreciable effect on the system’s behavior. This choice is also consistent with prior work~\cite{cuevas2022}.

It has been shown that the mean-field implementation of the Yard-Sale model leads to a final state in which a single agent accumulates all the wealth, regardless of the initial wealth or risk distributions~\cite{CAON2006, BOGHOSIAN2015, chorro2016simple, CARDOSO2021, CARDOSO2023}.

\subsection*{Social protection factor}
To prevent the undesirable outcome of wealth concentration, we employ a commonly used mechanism that favors the poorer agent in each transaction~\cite{SCAFETTA2002, Scafetta2004, Iglesias2004}. This mechanism introduces an asymmetry in the distribution of $\eta_{ij}$, implemented through the following expression for the probability $p$:
\begin{equation}
p = \frac{1}{2} + f \frac{|w_i - w_j|}{w_i + w_j},
\end{equation}
where $f$ is called the \textit{social protection factor} and takes values between $0$ (equal probability of winning for both agents) and $0.5$ (maximum bias in favor of the poorer agent). We treat $f$ as a key parameter of the system and analyze how changes in its value affect the resulting wealth distribution.

\subsection*{Growth and redistribution}

To explore how economic growth influences the wealth distribution, we adopt a generalization of the Yard-Sale model~\cite{liu2021}, in which the total wealth grows exponentially at a rate regulated by a parameter $\mu$. Thus, after a complete exchange step (consisting of $N$ Yard-Sale exchanges), an amount $\mu W(t)$ of wealth is added to the system, which is distributed among the agents according to

\begin{equation}\label{eq:redistribution}
w_i^* = w_i + \mu W \frac{w_i^{\lambda}}{\sum_{j=1}^{N} w_j^{\lambda}},
\end{equation}
where $\lambda \geq 0$, called the distribution parameter, controls which segment of the population benefits the most from additional wealth: if $\lambda < 1$, the distribution favors the poorer agents; if $\lambda > 1$, it favors the wealthier ones. After each wealth injection, the total wealth is renormalized to $W(t)=N$.

Since this study is primarily concerned with identifying mechanisms that prevent or reduce inequality, we focus on values of $\lambda \leq 1$, where redistribution favors the poorer agents. The case $\lambda > 1$, which amplifies wealth concentration, is beyond the scope of this work and is left for future investigation.

\subsection*{Algorithmic Framework and Numerical Implementation}

Given all these processes, a Monte Carlo step (MCS) of our simulation is defined as one execution of the following sequential steps:
\begin{enumerate}
    \item Choose $N$ random pairs $(i,j)$ and perform $N$ Yard-Sale exchanges given by rule~\ref{eq:ysrule} (agents can be selected more than once).
   
    \item Add new wealth $\mu W(t)$ and distribute it among the agents following Eq.~\ref{eq:redistribution}.
    \item Normalize the total wealth so that $W(t+1)=W(t)=N$.
\end{enumerate}

It is worth noting that our model reduces to the original model proposed in Ref.~\cite{liu2021} when $f = 0$, and recovers the Yard-Sale model with social protection but without wealth injection from~\cite{nener2021} when $\lambda = 1$. In this latter case, the redistribution process becomes neutral since the newly generated wealth is distributed proportionally to each agent's current wealth. From Eq.~\ref{eq:redistribution}, we have $w_i^* = (1 + \mu) w_i$, which implies that all agents' wealth grows exponentially at the same rate. This leads to a uniform scaling of the entire wealth distribution, preserving the relative differences among agents. Therefore, after normalizing the total wealth (e.g., fixing $W = N$), this global exponential growth factor cancels out, leaving the wealth distribution shape unchanged.

A commonly used statistical measure to quantify economic inequality in a system is the Gini index, defined for an ensemble of agents as

\begin{equation}
G = \frac{1}{2NW}\sum_{i=1}^{N} \sum_{j=1}^{N} |w_i - w_j|.
\end{equation}
The Gini index is zero when all agents have the same wealth and approaches one when a single agent holds all the wealth in the system. In our simulations, we also use the Gini index to determine when the system has reached a steady state. In particular, the temporal evolution of the Gini index serves as a robust indicator of the system’s behavior. As agents interact, the Gini index changes and eventually stabilizes at a constant value (within small fluctuations), indicating that the system has reached a stationary state. 

Unless otherwise stated, the results presented below were obtained via time averaging over a single realization of the dynamics. Once the system reached equilibrium (after $10^6$ MCS), we collected 1000 snapshots at intervals of 1000 MCS, spanning the range $[10^6, 2 \times 10^6]$. These snapshots were used to compute time-averaged wealth distributions and other quantities of interest. This approach is justified by the ergodic nature of the system under the mean-field interaction scheme, which ensures that long-time averages over a single realization are representative of ensemble behavior. To confirm this, we compared the time averages with ensemble averages computed over 100 independent realizations of the system, finding excellent agreement between the two, although these results are not shown here for the sake of brevity.

Based on the observed behavior of statistical fluctuations, the number of agents was fixed at \(N = 2500\), which ensures sufficiently low noise. The growth parameter was set at $\mu=0.001$, matching the value used in Section II of Ref.~\cite{liu2021}, which allows direct comparison with their results.

Finally, our simulations were performed by exploring a range of values for the redistribution parameter $\lambda$ and the social protection factor $f$, under two distinct risk schemes: fixed risk (constant $r_i$ for all agents) and random risk, where $r_i$ is drawn from a uniform distribution over the interval $[0,1]$. These schemes follow the frameworks proposed by Liu~\cite{liu2021} and Neñer~\cite{nener2021}, respectively, allowing us to analyze how different assumptions about risk heterogeneity affect the system’s behavior.

\section{Results}
In this section, we present the results obtained for our new version of the Yard-Sale model, which incorporates both the social protection factor and wealth injection and redistribution. In the first stage, we analyze the case in which the risk is constant for all agents, as proposed by Liu \textit{et al}.~\cite{liu2021}. In the second stage, we consider scenarios where the agents have randomly distributed risk factors, following the approach of Neñer and collaborators~\cite{nener2021}. 

\subsection{Constant Risk Case}

An initial characterization of the model involves analyzing the system’s behavior as a function of the two key parameters: the social protection factor $f$ and the redistribution factor $\lambda$. 

\begin{figure}[H]
\centering
    \includegraphics[width=0.95\columnwidth]{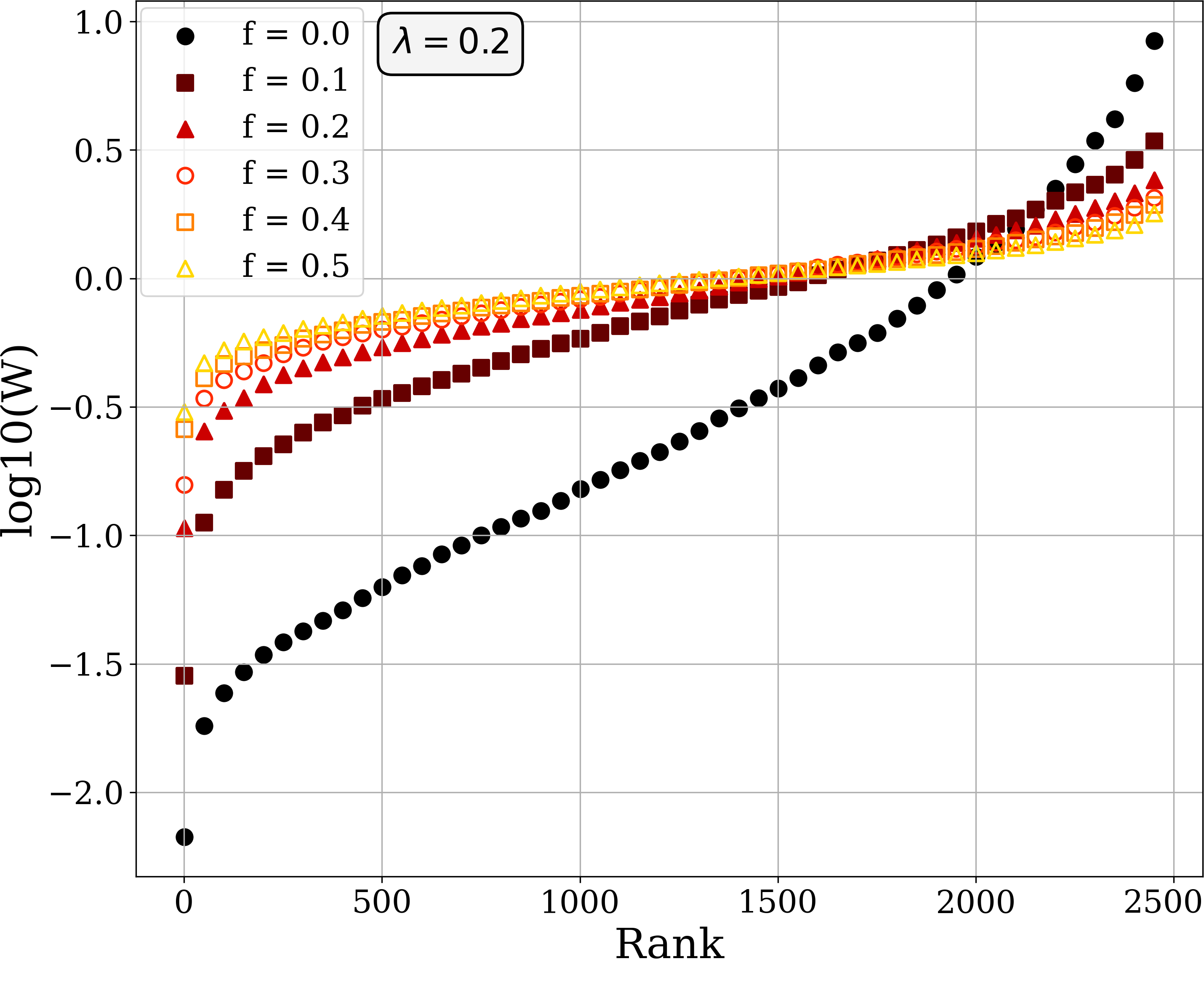}
    \caption{Wealth as a function of agent rank, for fixed risk $r_i = 0.1$ and $\lambda = 0.2$, shown for different values of the social protection factor $f$. One point is plotted every 50 agents out of a total population of 2500. Each curve corresponds to a single realization of the dynamics, taken in the stationary regime after $10^6$ MCS.}

\label{fig13}
\end{figure}

To enable direct comparison with the results reported by Liu \textit{et al}.~\cite{liu2021}, we begin by presenting the wealth distributions ordered by agent rank—a representation adopted in their work to visualize inequality patterns.
Figure~\ref{fig13} shows the rank-ordered wealth distributions for the case $\lambda = 0.2$ and several values of $f$. As $f$ increases, the curves flatten and the disparities at both ends become less pronounced, reflecting a more equitable distribution of wealth among agents. The case $f = 0$ closely matches the rank-ordered wealth distribution presented in Fig. 2 of Ref.~\cite{liu2021}, serving as a validation of our implementation.

An alternative way to examine how wealth is distributed among the population of agents is through the construction of wealth distribution curves, which are a standard tool in models of this kind for assessing inequality. These curves are typically derived from histograms which, although often noisy and sensitive to parameter variations, provide valuable qualitative insights into the system’s behavior.

\begin{figure}[H]
\centering
    \includegraphics[width=0.95\columnwidth]{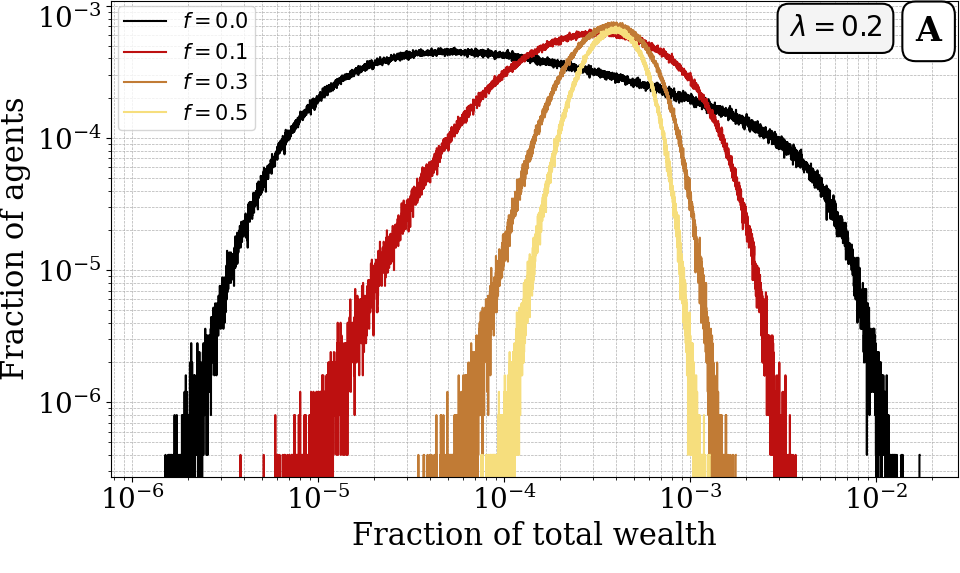}
    \includegraphics[width=0.95\columnwidth]{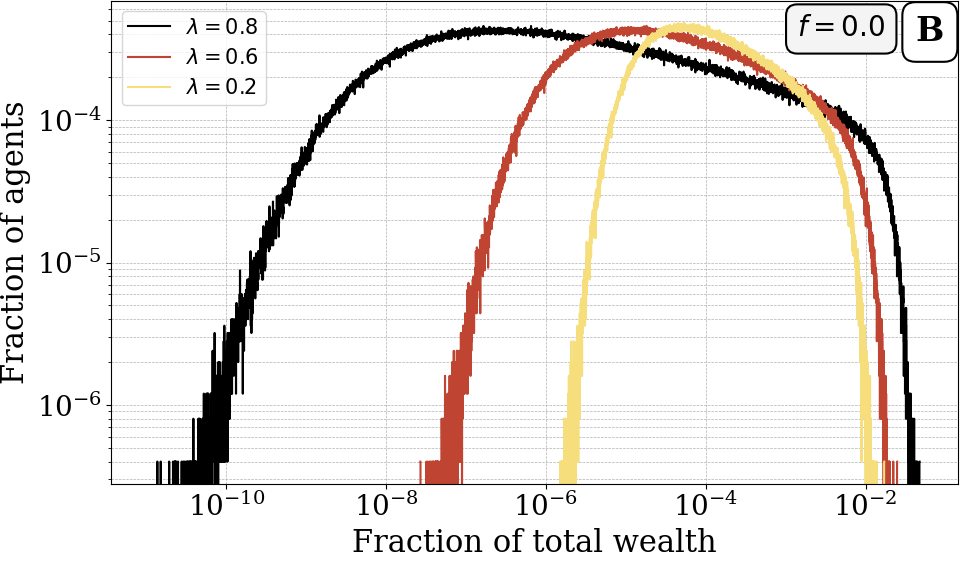}
    \includegraphics[width=0.95\columnwidth]{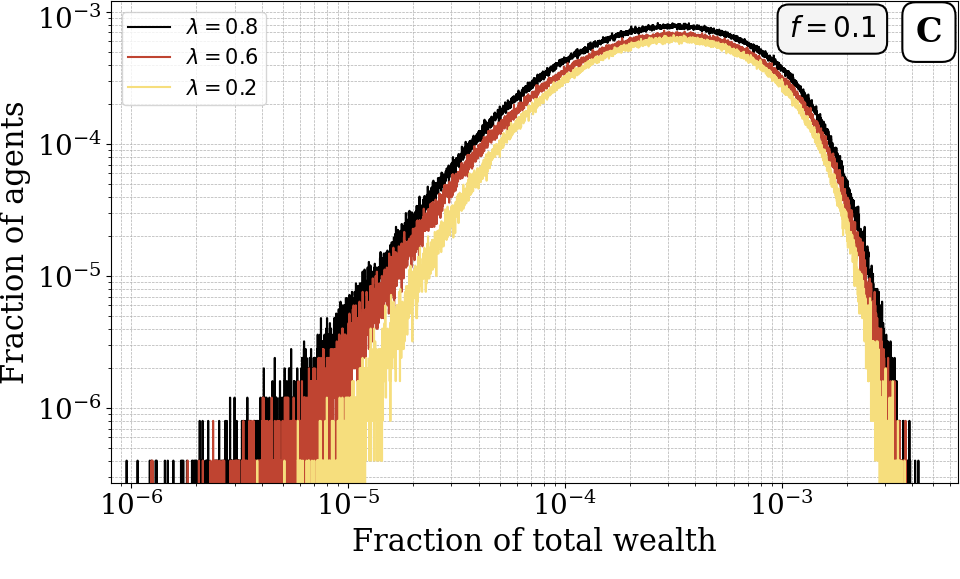}
    \caption{Wealth distributions at stationary state with logarithmic binning. (A) $\lambda = 0.2$ and several values of $f$. (B) $f = 0$ and several values of $\lambda$. (C) $f = 0.1$ and several values of $\lambda$. In all cases, $r_i = 0.1$ for all agents. Each histogram corresponds to a temporal average over 1000 configurations.}    
\label{w-rfijo}
\end{figure}

Figure~\ref{w-rfijo}A shows the resulting wealth distributions for different values of the social protection factor $f$, keeping $\lambda$ fixed. As $f$ increases, the distribution becomes more compressed: the lower tail (representing poorer agents) shifts to the right, while the upper tail (representing wealthier agents) moves to the left. This contraction reduces the wealth gap between rich and poor, promoting the emergence of a more pronounced middle-class peak. This result is consistent with the behavior observed in the individual wealth profiles shown as a function of rank, where increasing $f$ also led to a reduction in inequality.

To complement this analysis, we present the results when $f$ is held constant and $\lambda$ is varied. Figure~\ref{w-rfijo}B shows the system’s behavior in the absence of social protection ($f = 0$) for three different values of $\lambda$. We observe that decreasing $\lambda$ produces an effect qualitatively similar to increasing $f$: the wealth distribution becomes more egalitarian, with reduced disparities between the richest and poorest agents and a more prominent middle-class peak. However, the efficacy of $\lambda$ in reducing inequality is notably lower. This can be clearly seen by comparing the horizontal scales of panels A and B in Figure~\ref{w-rfijo}, which reveal that the overall wealth gap remains wider when only redistribution is applied. When $f = 0.1$ (Fig.~\ref{w-rfijo}C), the impact of $\lambda$ on the distribution becomes even weaker. Only slight shifts in the tails are observed, leading to a marginal narrowing of the curve. This highlights a marked difference in the influence of $f$ versus $\lambda$ on the shape of the wealth distribution

\begin{figure}[H]
    \centering
    \includegraphics[width=.95\columnwidth]{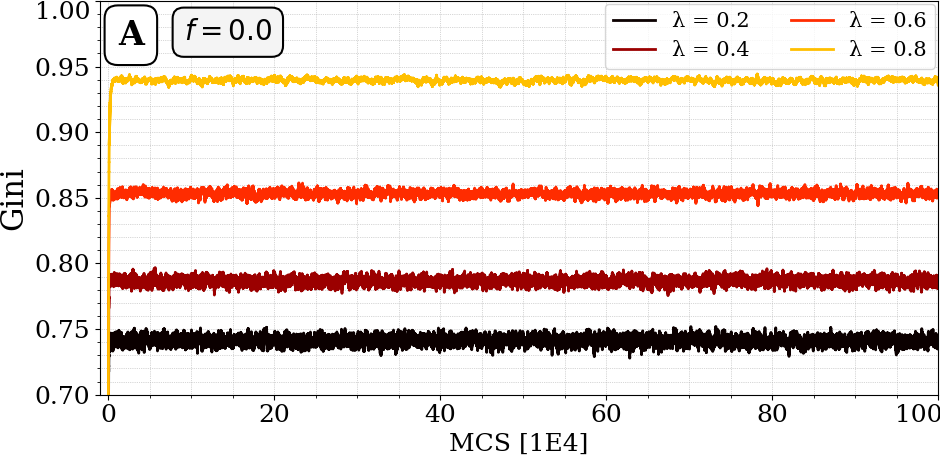}
     \includegraphics[width=.95\columnwidth]{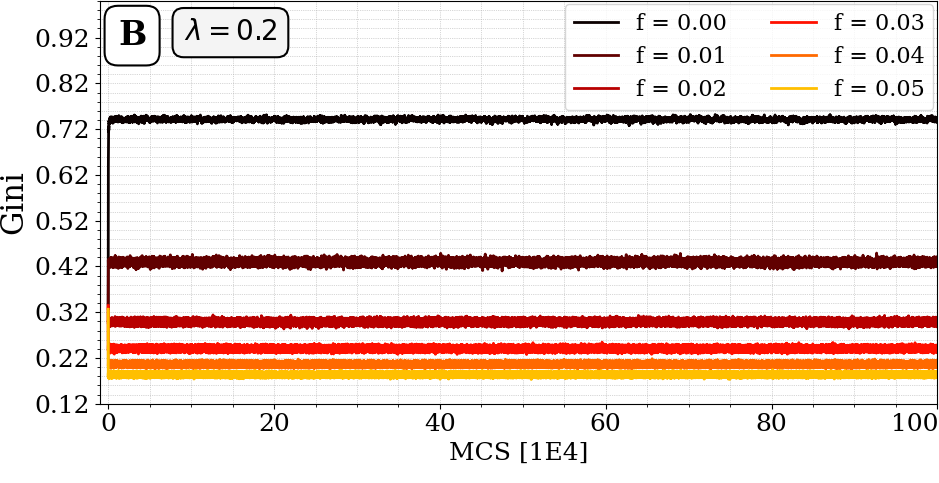}
    \caption{Gini coefficient as a function of Monte Carlo steps. (A) $f=0$ for different values of $\lambda$. (B) $\lambda=0.2$ for different values of the social protection factor $f$. In all cases, the risk factor is $r=0.1$ for all agents. Each curve corresponds to a single realization of the dynamics. }
       
\label{fig7}
\end{figure}

The previous analysis focused on the stationary wealth distributions, offering a snapshot of inequality at equilibrium. To complement this perspective, we now examine how the system evolves over time through the Gini index. Figure~\ref{fig7} shows its evolution for different values of $\lambda$ and $f$, assuming a constant risk factor of $r_i=0.1$ for all agents. The top panel corresponds to the case with $f=0$, which reproduces the model of Ref.~\cite{liu2021}, while the bottom panel displays the system’s behavior as a progressively higher social protection factor is introduced for a single value of $\lambda$. 

An initial analysis shows that, in all the cases, the temporal evolution of the complete model is consistent with that observed in previous versions, as the Gini index stabilizes within a few Monte Carlo steps, indicating the attainment of a steady state. 

From Fig.~\ref{fig7}A, we observe that the stationary value of the Gini coefficient decreases as $\lambda$ decreases, for a fixed value of $f = 0$. This behavior highlights the mechanism by which the wealth injection governed by $\lambda$ selectively benefits agents whose wealth has dropped below the minimum threshold, allowing them to re-enter the dynamics. As a result, the system recovers previously inactive agents and reduces overall inequality. Besides, although residual fluctuations tend to increase with $\lambda$, the chosen system size of $N=2500$ is sufficiently large to ensure well-defined stationary states throughout our simulations.
Moreover, in Fig.~\ref{fig7}B we analyze the system’s behavior for several values of $f > 0$ with a fixed $\lambda = 0.2$. The qualitative behavior is consistent with that observed in the basic model without wealth injection: as $f$ increases, the Gini coefficient decreases. 

\begin{figure}[H]
\centering
    \includegraphics[width=0.9\columnwidth]{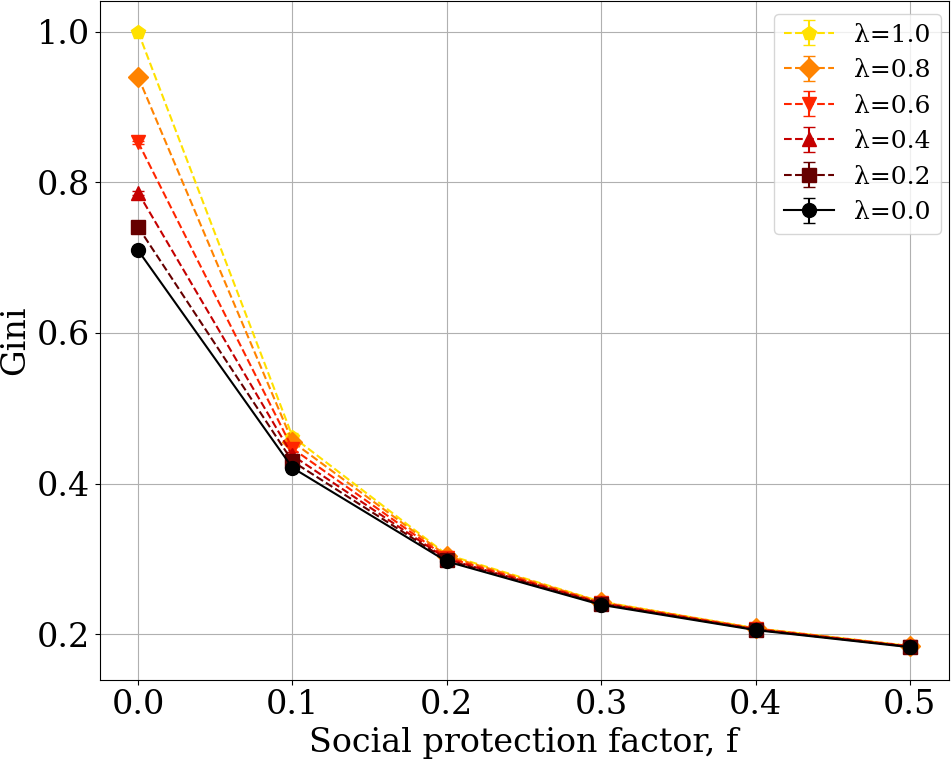}
    \caption{Gini coefficient as a function of $f$, with constant risk set at $r_i=0.1$, for different values of $\lambda$. Each point is a temporal average over 1000 configurations. Error bars are included but lie within the symbol size.
      }
\label{gini-rfijo}
\end{figure}

To illustrate the combined effect of the social protection and redistribution parameters, Fig.~\ref{gini-rfijo} shows the stationary Gini coefficient as a function of $f$ for different values of $\lambda$. A first observation is that both increasing the social protection factor $f$ and decreasing the redistribution threshold $\lambda$ lead to a reduction in the Gini index, as expected.
A closer examination reveals that for values of $f \gtrsim 0.2$, the curves corresponding to different values of $\lambda$ monotonically converge toward the one obtained for $\lambda = 0$. This collapse suggests that, beyond this threshold, the effect of $\lambda$ becomes negligible. In other words, once a sufficiently strong level of social protection is introduced, the redistribution mechanism governed by $\lambda$ no longer significantly influences the wealth distribution. This indicates that the social protection factor $f$ plays a substantially more dominant role in shaping the system’s long-term behavior, whereas $\lambda$ primarily acts as a reintegration mechanism for agents that fall below the threshold and are temporarily excluded after each Monte Carlo step.

\subsection{Random Risk Case}

We now consider a scenario in which agents face heterogeneous risk levels, randomly assigned from a uniform distribution over the interval $[0, 1]$, following the approach proposed in Ref.~\cite{nener2021}. As anticipated, introducing risk variability induces qualitative changes in the system’s behavior compared to the fixed-risk case. 
In what follows, we analyze how this heterogeneity affects the wealth distribution, the stationary Gini coefficient, and the effectiveness of the redistribution and social protection mechanisms.

\begin{figure}[H]
    \centering
    \includegraphics[width=0.88\columnwidth]{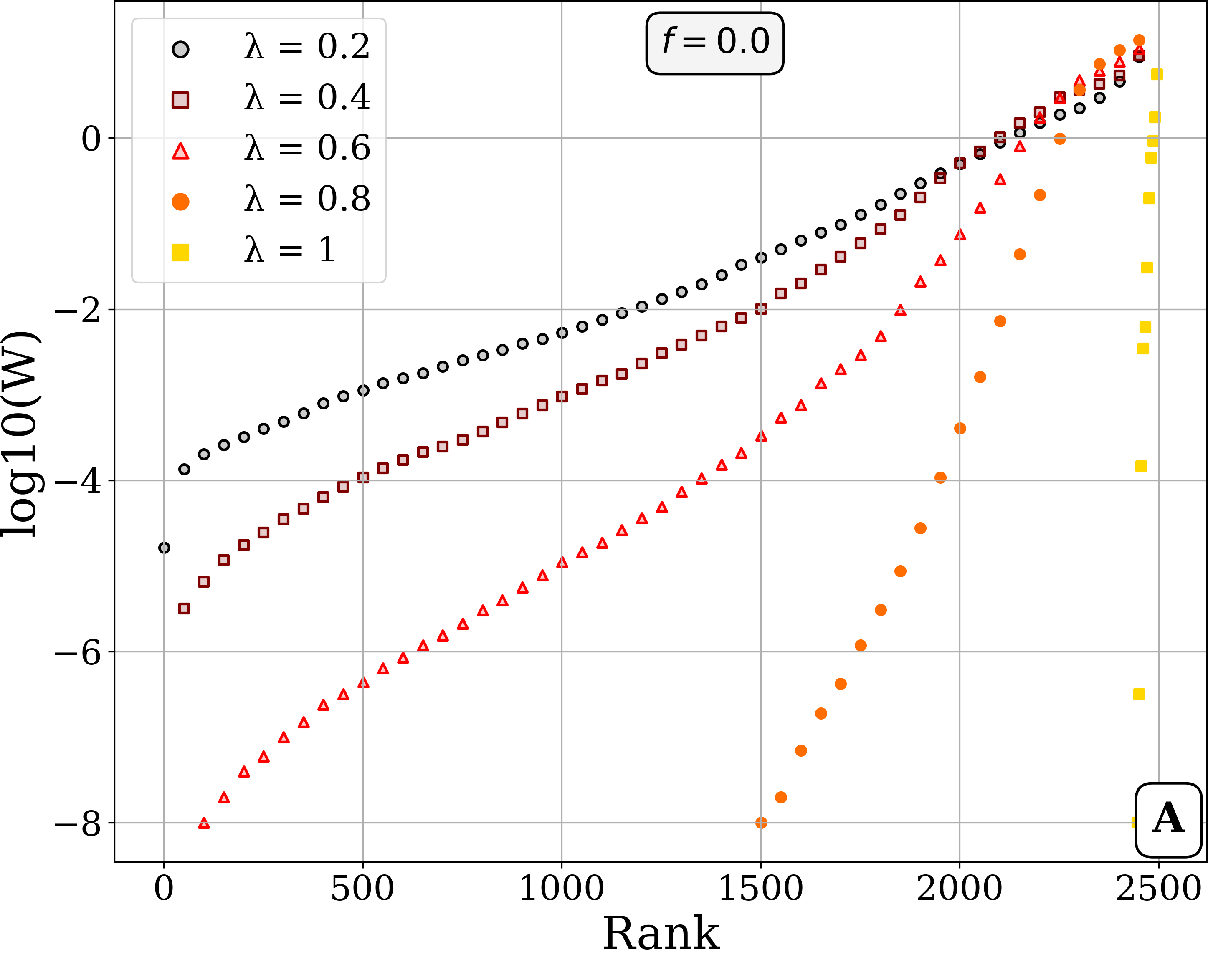}
    \includegraphics[width=0.88\columnwidth]{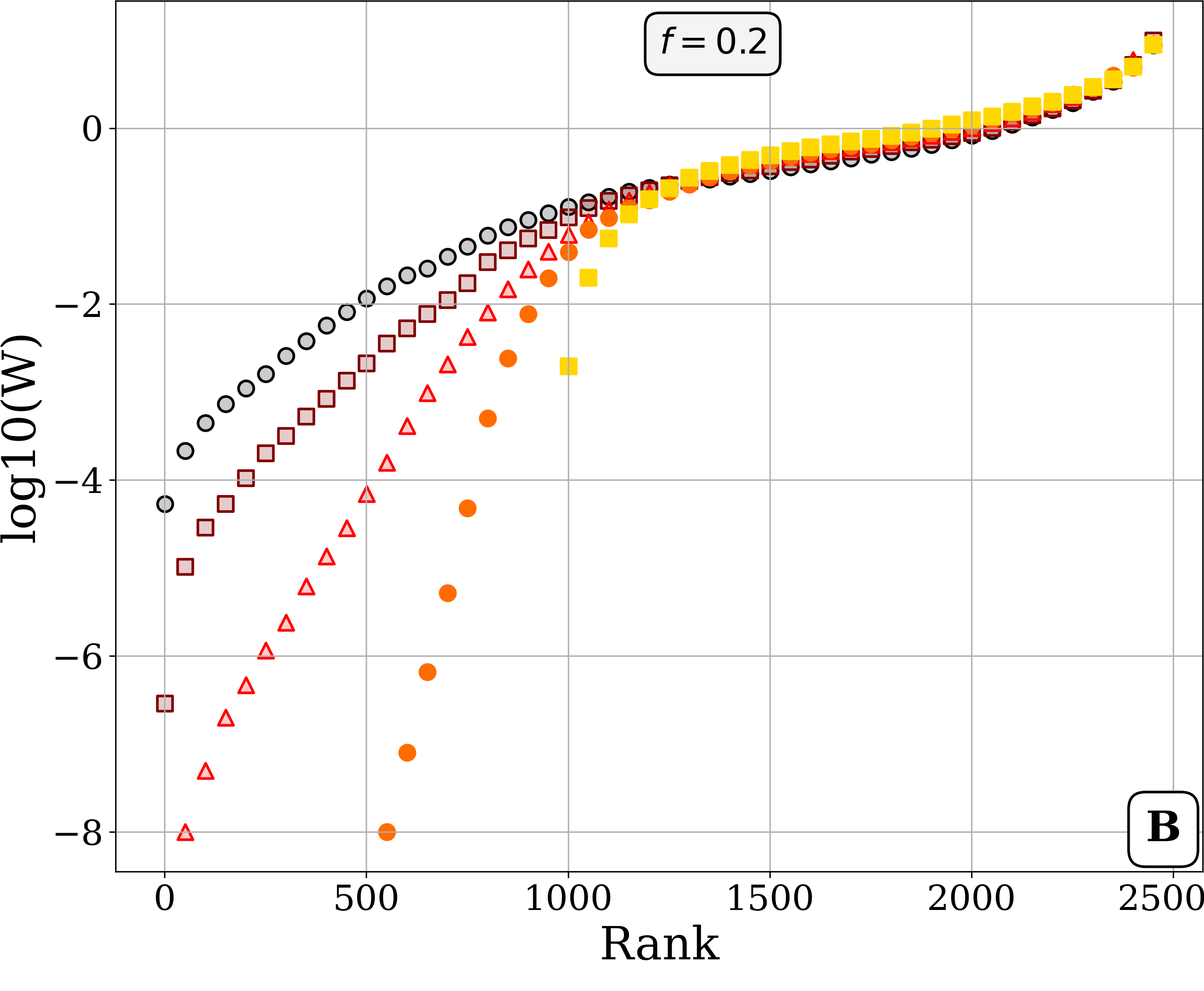}
     \includegraphics[width=0.88\columnwidth]{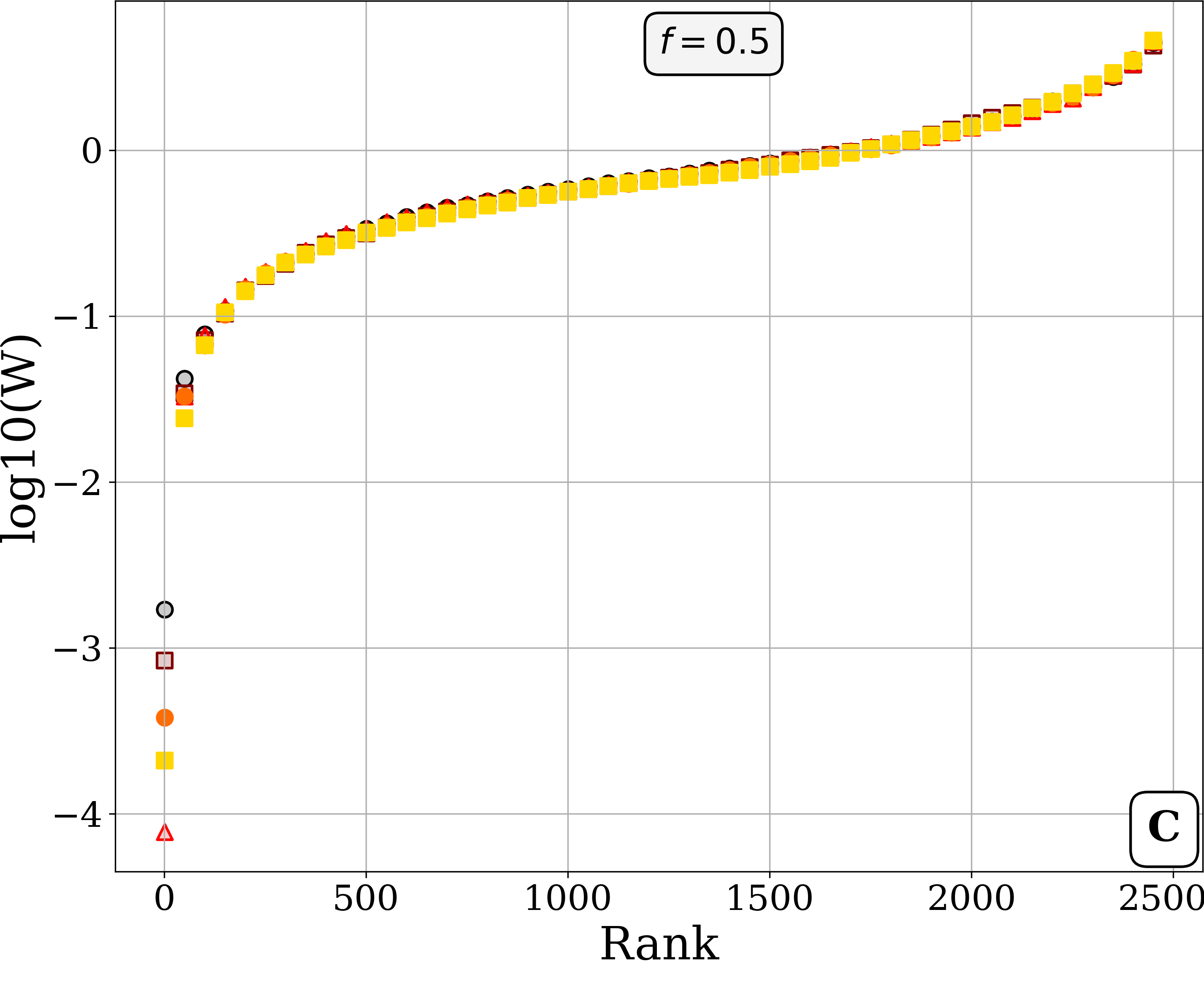}
    \caption{Wealth as a function of agent rank, for agents with risk values uniformly distributed in the range [0, 1]. (A) $f = 0$ and several values of $\lambda$. (B) $f = 0.2$ and several values of $\lambda$. (C) $f = 0.5$ and several values of $\lambda$. One data point is plotted for every 50 agents, out of a total population of 2500. Each curve represents a single realization, taken in the stationary regime after $10^6$ MCS.
    }
\label{w-rank3}
\end{figure}

As a first step in the characterization of the variable-risk scenario, Fig.~\ref{w-rank3} presents the rank-ordered wealth distributions for three different values of $f$ and several values of $\lambda$.

For the case of $f=0$, as shown in Fig. \ref{w-rank3}A, the redistribution parameter $\lambda$ is not sufficient to reincorporate all the agents excluded from the system in the previous Monte Carlo step. This situation persists until $\lambda$ becomes less than approximately 0.6, at which point it effectively compensates for the losses and ensures that all agents rejoin the system’s dynamics. As the value of $f$ increases (Fig. \ref{w-rank3}B), the social protection mechanism prevents the emergence of agents with negligible wealth, reducing the required value of $\lambda$ for full reincorporation. In the limiting case where $f=0.5$ presented in Fig. \ref{w-rank3}C, no agents are lost from the system, and the wealth distribution becomes independent of $\lambda$. This convergence of the wealth distribution for sufficiently high values of $f$ further confirms the dominant role of social protection in ensuring equitable outcomes, rendering the redistribution parameter $\lambda$ largely irrelevant beyond this threshold.

\begin{figure}[H]
\centering
    \includegraphics[width=0.9\columnwidth]{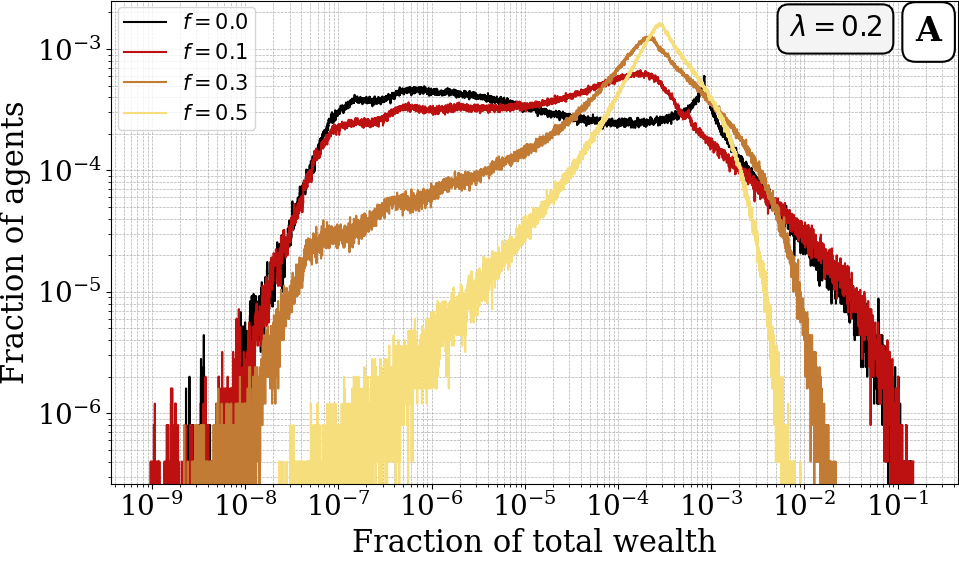}
    \includegraphics[width=0.9\columnwidth]{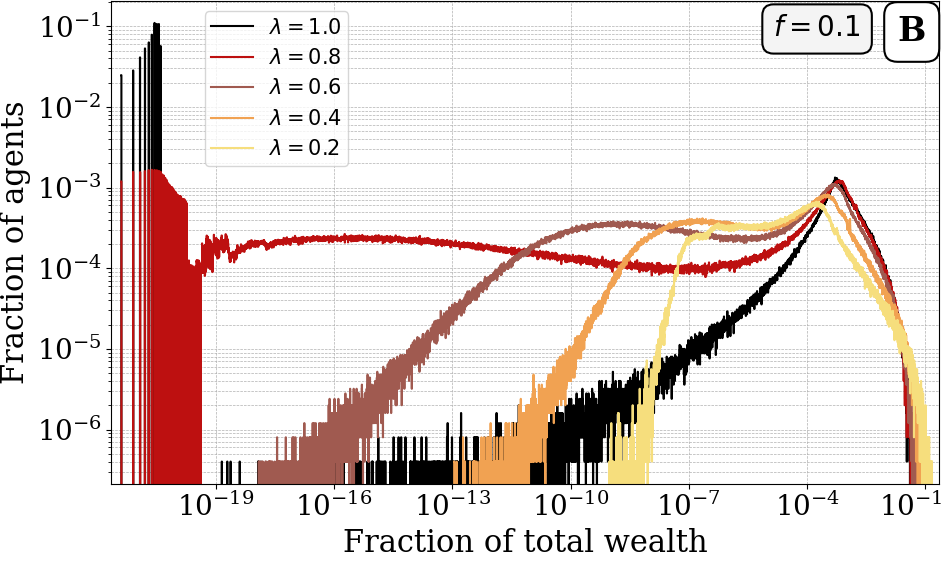}
    \caption{Stationary wealth distributions with logarithmic binning for agents with risk values uniformly distributed in the range [0, 1]. (A) $\lambda = 0.2$ and several values of $f$. (B) $f =0.1$ and several values of $\lambda$. Each histogram corresponds to a temporal average over 1000 configurations.}
\label{w-rvar}
\end{figure}

Building on the insights gained from the rank-ordered wealth distributions, we now turn to the analysis of the full wealth distributions. In Fig.~\ref{w-rvar}, we present the wealth distributions at the stationary state for various values of the social protection and redistribution factors.

 A comparison with Fig.~\ref{w-rfijo} reveals both similarities and differences. Among the similarities, we observe that increasing $f$ (as shown in panel (A)) leads to more equitable distributions, as the ends of the curves draw closer together, reducing the gap between rich and poor. However, there is a substantial difference in the shape of the curves: whereas in the fixed-risk case the curves were always concave, here the functional form is more complex. Specifically, a local minimum appears in the region of intermediate wealth, whose depth and position depends on the value of $f$.

This nontrivial behavior is also observed in Fig.~\ref{w-rvar}B, where we explore the effect of varying $\lambda$ for a fixed social protection factor $f = 0.1$. Notably, the system now exhibits a marked sensitivity to changes in $\lambda$, a feature absent in the fixed-risk case (compare with Fig.~\ref{w-rfijo}B). 
The curve corresponding to $\lambda = 1$ presents a pronounced accumulation of agents with wealth below $w_{\min}$. These agents, effectively excluded from the dynamics in the basic Yard-Sale model, form a sharp peak below $10^{-17}$ in the wealth histogram. Since the figure displays the fraction of total wealth, the plotted values are normalized by the number of agents (2500), causing this peak to appear below $10^{-19}$.

As $\lambda$ decreases, redistribution begins to inject wealth into the poorest sector, progressively reincorporating these inactive agents into the dynamics. This is reflected in the emergence of a long left tail in the distribution for $\lambda = 0.8$. For lower values of $\lambda$, this effect intensifies: more agents re-enter the system, and they do so with greater accumulated wealth. Consequently, the tail shortens and shifts rightward, as seen in the case $\lambda = 0.2$. This process can be interpreted as a gradual reintegration of low-wealth individuals into the economy. This finding underscores how heterogeneity in individual risk can fundamentally alter the system’s response to redistribution policies.

\begin{figure}[H]
    \centering
    \includegraphics[width=0.9\columnwidth]{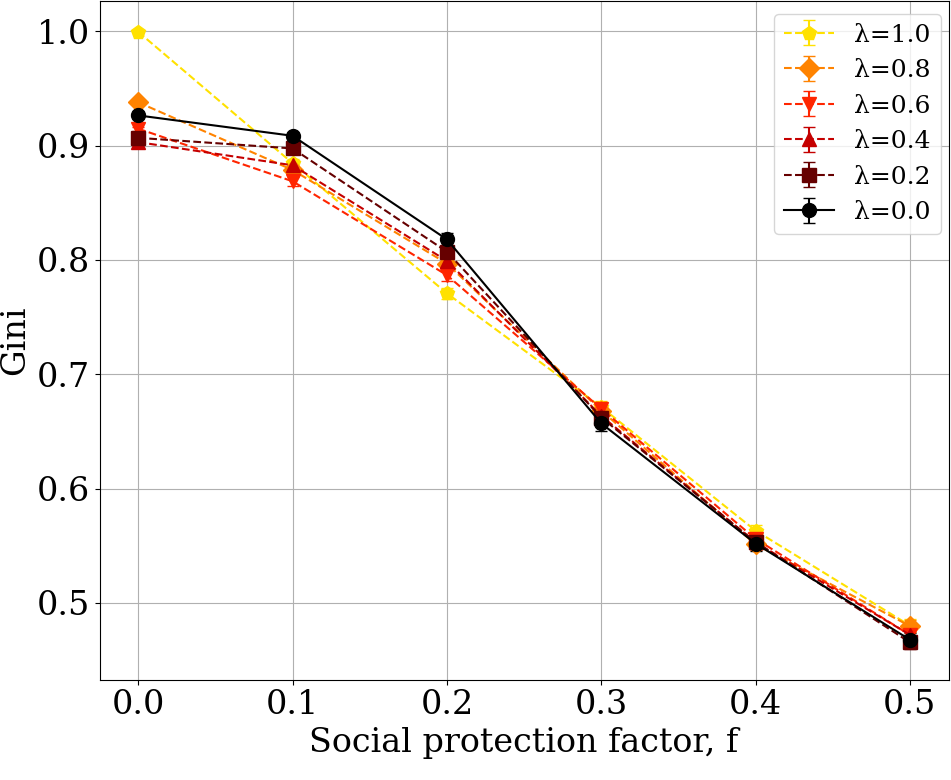}
    \caption{Gini coefficient as a function of the social protection factor $f$, for agents with risk values uniformly distributed in the range [0, 1], for different values of the redistribution parameter $\lambda$. Each point is an average over 1000 configurations. Error bars are included but lie within the symbol size.}
\label{gini-rvar}
\end{figure}

Another way of highlighting the differences between the two risk scenarios is by examining the stationary Gini index as a function of $f$ for various values of $\lambda$, as shown in Fig.~\ref{gini-rvar}. 
In the case where the individual risk factor $r_i$ is drawn from a uniform distribution, we observe a qualitatively similar pattern to the fixed-risk scenario presented in Fig.~\ref{gini-rfijo}: the Gini index as a function of $f$ shows that, for $f \gtrsim 0.2$, the curves corresponding to different values of $\lambda$ again collapse onto a single curve. This convergence suggests that, beyond this threshold, $\lambda$ ceases to have a significant impact on the system's inequality levels. 

However, marked differences emerge for smaller values of $f$. In contrast to the fixed-risk case, where the curves display a monotonic dependence on $f$ and $\lambda$, the randomly distributed risk case exhibits more complex patterns. Specifically, for any given $f$ the Gini values do not exhibit a consistent monotonic dependence on $\lambda$. This suggests a richer interplay between risk heterogeneity and redistribution dynamics in the low-social-protection regime, which we explore further below.

\begin{figure}[H]
    \centering
    \includegraphics[width=.95\columnwidth]{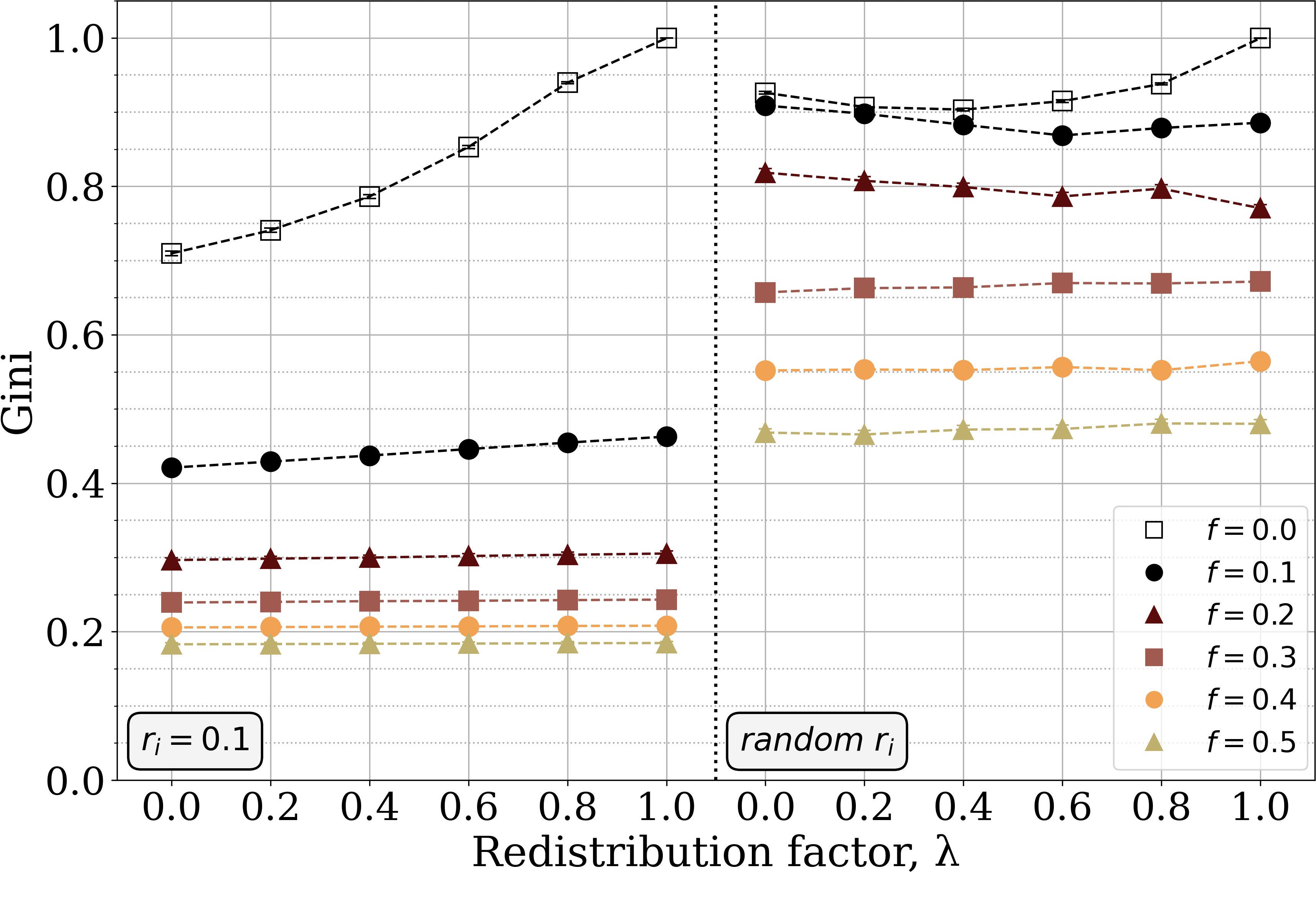}
    \caption{Gini coefficient as a function of the redistribution parameter $\lambda$, for different values of the social protection factor $f$. The left panel corresponds to agents with fixed risk $r_i = 0.1$, while the right panel corresponds to agents with risk values uniformly distributed in the range [0, 1]. Each point is an average over 1000 configurations. Error bars are included but lie within the symbol size.}
\label{gini_lambda_final}
\end{figure}

A useful way to understand this non-monotonic behavior is by integrating the curves from both cases into a single figure, allowing for a direct comparison across the full parameter space. This is shown in Fig.~\ref{gini_lambda_final}, which summarizes our main results and compares the two risk scenarios by presenting the stationary Gini index as a function of $\lambda$ for various values of $f$, considering both fixed and randomly distributed risks. In both cases, for $f \gtrsim 0.2$, the Gini index appears largely independent of $\lambda$, in agreement with the trends shown in Figs.~\ref{gini-rfijo} and~\ref{gini-rvar}. In contrast, the behavior for $f=0$ and $f=0.1$ differ between the two risk cases: for fixed risk, the Gini index increases with $\lambda$, consistent with Fig.~\ref{gini-rfijo}, whereas for random risk, the curves are non-monotonic and display a minimum. 
This difference arises from the interplay between risk heterogeneity and the redistribution mechanism defined by Eq.~\ref{eq:redistribution}. Although a detailed analysis of this interaction lies beyond the scope of the present work, it suggests a promising avenue for future research.

Moreover, Fig.~\ref{gini_lambda_final} also reveals that the case of fixed risks $r_i=0.1$ tends to result in lower Gini coefficients than the case of randomly assigned risks, across all values of $\lambda$ and $f$. This suggests that when all agents exhibit the same (and low) risk, the system is more effective at reducing inequality. The presence of heterogeneous risk preferences introduces additional variability that seems to hinder the equalizing effects of redistribution and social protection mechanisms. This phenomenon has been studied in~\cite{nener2021}, where the authors found that high-risk agents inevitably lose all their wealth, thereby increasing inequality. Our results suggest that this mechanism also operates in the extended model considered here.

\section{Discussion}

In this work, we have explored an extended version of the Yard-Sale model, incorporating a social protection factor and an exponential growth followed by a redistribution mechanism to mitigate wealth inequality. Our results show that these two mechanisms have a significant impact on the stationary wealth distribution of the system.

We found that the social protection factor $f$ plays a dominant role in shaping the system’s behavior. For sufficiently large values of $f$, the Gini coefficient reaches similar stationary values regardless of the redistribution parameter $\lambda$, indicating that the wealth distribution becomes largely insensitive to $\lambda$. In this regime, redistribution primarily serves to reincorporate agents whose wealth dropped below $w_{\text{min}}$ in the previous Monte Carlo step, rather than significantly altering inequality. Our findings point to the existence of a threshold around $f \gtrsim 0.2$, beyond which the protective effect becomes strong enough to suppress the impact of $\lambda$. Although no sharp transition is observed, this threshold marks a qualitative change in the system's response.

An additional key finding of this work is the strong dependence of the results on the underlying risk distribution. When the risk factor $r_i$ is homogeneous, \textit{i.e.}, identical for all agents, the system exhibits a monotonic relationship between the Gini coefficient and the parameters \(\lambda\) and $f$. In contrast, when the risk values are randomly assigned to agents, the Gini coefficient displays a non-monotonic dependence on \(\lambda\), with a pronounced minimum emerging for low values of $f$. This unexpected behavior reveals a subtle interaction between the redistribution dynamics and agent heterogeneity that is absent in the homogeneous case. These differences highlight the importance of considering agent-level variability in risk attitudes when modeling wealth exchange dynamics.

Several open questions remain. First, it would be valuable to extend the analysis by exploring different fixed values of $r_i$, as this study only considered a relatively low value ($r_i=0.1$).  

Furthermore, previous studies suggest that stationary wealth distributions may depend on the growth parameter $\mu$, which was kept fixed throughout this work. Exploring the effects of varying $\mu$ in our model could expand the understanding of the underlying dynamics and reveal whether the observed inequality patterns persist across different growth regimes.

Additionally, in the case of randomly distributed risk, a deeper understanding of the non-monotonic dependence of the Gini index on $\lambda$ is needed, as such behavior does not appear when risk is fixed. As suggested in Ref.~\cite{nener2021}, the system may exhibit a critical risk threshold above which agents tend to lose all their wealth. This threshold depends on $f$ and allows for a classification of strategies into potential winners and losers. The introduction of redistribution could interact non-trivially with this mechanism: for intermediate values of $\lambda$, differentiation between strategies might be reinforced, temporarily increasing inequality, while stronger or weaker redistribution would suppress this effect. This could explain the emergence of a minimum in the Gini index.

Overall, our study provides quantitative evidence that social protection and economic growth with redistribution mechanisms significantly enhance equality and can counterbalance the inherent tendencies of the Yard-Sale model towards wealth condensation. Another important point is that both the shape of the wealth distribution and the resulting inequality levels are strongly influenced by the underlying distribution of individual risk, highlighting the importance of considering agent heterogeneity in policy design. These results encourage further research into how social protection, redistribution, and individual heterogeneity interact in more realistic economic environments.

\section{Acknowledgments}
The authors thank Ignacio Cortés, from Centro Atómico Bariloche (CNEA, Argentina) for fruitful discussions.


\end{document}